\documentclass[runningheads]{llncs}
\usepackage[T1]{fontenc}
\usepackage{booktabs}
\usepackage{graphicx}
\usepackage{color}
\usepackage{url}

\begin{document}
\sloppy

\title{Securing the Web: Analysis of HTTP Security Headers in Popular Global Websites}

\titlerunning{Analysis of HTTP Security Headers in Popular Global Websites}

\author{Urvashi Kishnani\orcidID{0000-0001-6389-5508} \and
Sanchari Das\orcidID{0000-0003-1299-7867}}

\authorrunning{Kishnani and Das}

\institute{University of Denver, Denver CO 80208\\
\email{\{Urvashi.Kishnani,Sanchari.Das\}@du.edu}}
% \and
% George Mason University, Fairfax, Virginia\\
% \email{sdas35@gmu.edu}}

% \author{Urvashi Kishnani \inst{1}\orcidID{0000-0001-6389-5508} \and
% Sanchari Das\inst{1,2}\orcidID{0000-0003-1299-7867}}

% \authorrunning{Kishnani and Das}

% \institute{University of Denver, Denver CO 80208
% \email{\{Urvashi.Kishnani,Sanchari.Das\}@du.edu}\\
% \and
% George Mason University, Fairfax, Virginia\\
% \email{sdas35@gmu.edu}}
%
\maketitle              
\begin{abstract}
The surge in website attacks, including Denial of Service (DoS), Cross-Site Scripting (XSS), and Clickjacking, underscores the critical need for robust HTTPS implementation—a practice that, alarmingly, remains inadequately adopted. Regarding this, we analyzed HTTP security headers across $N=3,195$ globally popular websites. Initially, we employed automated categorization using Google NLP to organize these websites into functional categories and validated this categorization through manual verification using Symantec Sitereview. Subsequently, we assessed HTTPS implementation across these websites by analyzing security factors, including compliance with HTTP Strict Transport Security (HSTS) policies, Certificate Pinning practices, and other security postures using the Mozilla Observatory. Our analysis revealed over half of the websites examined ($55.66\%$) received a dismal security grade of `F' and most websites scored low for various metrics, which is indicative of weak HTTP header implementation. These low scores expose multiple issues such as weak implementation of Content Security Policies (CSP), neglect of HSTS guidelines, and insufficient application of Subresource Integrity (SRI). Alarmingly, healthcare websites ($n=59$) are particularly concerning; despite being entrusted with sensitive patient data and obligations to comply with data regulations, these sites recorded the lowest average score ($18.14$). We conclude by recommending that developers should prioritize secure redirection strategies and use implementation ease as a guide when deciding where to focus their development efforts.
\end{abstract}

\section{Introduction}
Websites are essential in the digital age, offering services across e-commerce, banking, healthcare, and more~\cite{surani2022understanding,debnath2019study,debnath2020studies,gopavaram2019iotmarketplace}. They, along with mobile apps, have seen a significant increase in traffic~\cite{kishnani2023assessing,neupane2022data,tazi2023accessibility,dev2018privacy,das2019towards}. For example, in November $2022$, YouTube and Facebook logged 74.8 billion and $10.7$ billion monthly views, respectively~\cite{bianchi_2023,das2024design,momenzadeh2020bayesian}. The most popularly used secure protocol for data exchange between websites and clients is the HyperText Transfer Protocol with the added Transport Layer Security (TLS) giving HTTPS~\cite{callegati2009man,das2020user}. The research community has long recognized the importance of securing websites and has developed various tools, techniques, and best practices to improve web security~\cite{shi2010analysis,das2020risk,noman2019techies,das2019privacy,surani2023security1,noah2022privacy}. However, the implementation of these security practices remains inconsistent across different website categories and industries~\cite{mendoza2018uncovering,tazi2022sok,kishnani2022privacy,fernandes22you}. Moreover, research indicates that the complexity of the web ecosystem can magnify the impact of even a few exploitable HTTPS vulnerabilities~\cite{hadan2019making,surani2023security,calzavara2019postcards,walsh2021my,lichlyter2024understanding}. The security of websites, including HTTP header implementation, is even more critical for websites involving financial transactions. This is particularly important for popular websites, as increased traffic correlates with more transactions. Thus, our study aims to answer the following research questions:

\begin{itemize}
\item \textbf{RQ1:} To what extent are the implementation and adherence to HTTP security headers, notably redirection protocols, the utilization of secure cookies, and Content Security Policy (CSP) directives, consistently observed as points of vulnerability or misconfiguration among widely-visited websites?
\item \textbf{RQ2:} How does the implementation and compliance of multiple HTTP security headers vary across diverse website categories, including but not limited to Computers \& Electronics, Finance, and Health? And how can these variations be ranked or categorized based on their robustness or potential vulnerabilities?
\end{itemize}

We collected a list of the $10,000$ most popular websites and categorized them into $27$ categories like Shopping, Games, etc. using Natural Language Processing (NLP). This categorization assists in linking these website categories to user-friendly functional groups. Next, we conducted a security evaluation of $3,195$ categorized websites and analyzed their aggregated, category-wise, and individual security metric performance using Mozilla Observatory~\cite{observatory}. 

 \begin{itemize}
\item Our study evaluates website security through the lens of popular website categories, an understudied area. This categorization helps us analyze differences in security measures among various types of websites correlated with their popularity and categorization, allowing us to identify specific vulnerabilities and tailor security improvements accordingly. Moreover, although security headers are typically among the initial aspects of security to be managed and tested in the web security community, the reality on a global scale appears to be concerning. 
\item Our results emphasizes the need for enhanced security measures on websites categorized as health, sports, and news, which have lower overall security rankings. Alarmingly, healthcare websites have the lowest security ratings, posing risks to users and potential non-compliance with global healthcare regulations. Furthermore, computer security websites scored poorly, with an average score of $31.89$, which is concerning as well. In addition, $55.66\%$ of the examined websites received a low security grade of `F', indicating weak HTTP headers implementation that may leave these websites vulnerable various cyber threats, including XSS, man-in-the-middle (MiTM) attacks, clickjacking, and data breaches. 
\end{itemize}

\section{Related Works}
\label{sec:related_works}
\subsection{Website Classification \& Categorization} Prior studies have utilized URLs as the primary input for categorizing websites using the Naive Bayes approach~\cite{rajalakshmi2011naive,das2023review}. URL analysis has also been used to filter malicious sites by examining lexical features with a Naive Bayes classifier~\cite{aldwairi2012malurls}. Various works focus on detecting phishing websites through classification techniques~\cite{das2019all,gopavaram2021cross,tally2023mid,noah2022phishercop,unchit2020quantifying,das2022evaluating,tally2023tips}. Shabudin et al. proposed feature selection techniques to detect phishing sites~\cite{shabudin2020feature}. Content-based website categorization faces challenges such as large data volumes, effective scraping, and robust algorithm training~\cite{bruni2020website}. Bruni and Bianchi used web scraping and OCR for e-commerce website categorization, though their approach was limited~\cite{bruni2020website}. Our research categorizes websites based on their function, analyzing homepage content rather than URLs to create functional groups, avoiding binary phishing detection.

\subsection{Website Security} Effective website security evaluation tools should ensure compatibility, crawlability, vulnerability testing, detailed reporting, and usability with up-to-date databases and flexible interfaces (GUI or CLI)\cite{shi2010analysis,jiang2011security}. Mozilla Observatory meets these requirements and has been used to measure HTTPS protocol adoption\cite{felt2017measuring} and analyze HTTPS headers on a large scale~\cite{lavrenovs2018http}. Fonseca et al. evaluated website security using SQL injection and XSS testing~\cite{fonseca2013evaluation}, while Johns et al. developed XSSDS for detecting XSS attacks~\cite{johns2008xssds}. Cernica et al. assessed website security via WordPress backup plugins~\cite{cernica2019security}, and Szydlowski et al. highlighted server-side solutions for mitigating client-side attacks like Trojan Horses~\cite{szydlowski2007secure}. Jammalamadaka et al. introduced “Delegate,” a proxy-based architecture for secure web access~\cite{jammalamadaka2006delegate}. Al et al. used w3af~\cite{w3af} and Skipfish~\cite{skipfish} for a security evaluation of Saudi Arabian websites~\cite{al2015security}. Our study adopts a similar approach to analyze HTTP security headers, with the added aspect of website categorization using NLP. Felt et al.\cite{felt2017measuring} and Lavrenovs et al.\cite{lavrenovs2018http} used Mozilla Observatory to track HTTPS header adoption. Gadient et al. found a lack of secure HTTP headers in mobile app URLs~\cite{gadient2021security}, and Mendoza et al. revealed inconsistencies in HTTP security across various web and mobile categories~\cite{mendoza2018uncovering}. Stock et al. analyzed web security mechanisms in modern standards, suggesting areas for improvement~\cite{stock2014state}, while Chen et al. identified gaps in developers' understanding of CORS~\cite{chen2018we}. Meiser et al. highlighted how domain relaxation via CORS adds security risks when subdomains are hosted by third parties~\cite{meiser2021careful}.

\section{Method}
We developed a method to collect, categorize, and assess top websites' security, using Python for web scraping, Google NLP for categorization~\cite{website_categorization}, and Mozilla Observatory for HTTPS header evaluation, across four main stages.

\subsection{Resource Gathering}
To form our initial data corpus, we first gathered the top websites using Tranco's list, which provides the most popular one million domains~\cite{pochat2018tranco}. The data, collected on February 19, 2023~\footnote{Available at~\url{https://tranco-list.eu/list/3VJNL}}, is based on aggregated rankings from various tools, such as Cisco Umbrella~\cite{shah2017cisco}, Majestic~\cite{majestic}, Quantcast~\cite{quancast}, and Farsight~\cite{farsight}. By considering the website ranking over the last $30$ days, the list minimizes the likelihood of manipulation. For our study, we selected the top $10,000$ websites from the Tranco list as the basis for website categorization.

\subsection{Preparation of Dataset}

\subsubsection{Website Scraping and Content Collection}
To collect the text data from a website, we used the \texttt{cloudscraper 1.2.69}~\cite{cloudscraper} Python module. This module is built to bypass Cloudflare's anti-bot page, which allows large scale crawling or scraping of websites without being blocked by Cloudflare. Websites not using Cloudflare will be handled as is. The module uses JavaScript to impersonate a regular web browser. The request is made to each website using the following \texttt{user-agent} header: \texttt{Mozilla/5.0 (Windows NT 10.0; Win64; x64) AppleWebKit/537.36 (KHTML, like Gecko) Chrome/110.0.0.0 Safari/537.36}. Once the content is scraped from the website, we used the Beautiful Soup Python module~\texttt{beautifulsoup4 4.11.2}~\cite{beautiful_soup} to find HTML elements of interest using the appropriate tags: website's title using \texttt{title} tag, website's description \texttt{description} tag, website's content headers using \texttt{h1, h2, h3} tags, and website's content paragraph using \texttt{p} tag. This content was scraped from the landing page of the website, and not from any internally linked pages. The content from each tag was collected into one string which was then used for translation. We obtained content for $7,410$ out of the $10,000$ websites.

\subsubsection{Translating Content}
Using \texttt{deep-translator 1.10.1}~\cite{deep_translator} Python module, we translated the collected content into English. This module automates bulk translations and provides support for multi-language translation through various translation libraries, such as Google Translate, Microsoft Translate, and Libre Translator. We employed the Google Translate feature, which allows for only $1,000$ characters to be translated in a single document. Websites are recommended to have the length of their title of about six to eight words and their description of no more than $250$ characters~\cite{wilson2006search}. Thus, we think using the first $1,000$ characters of our scraped content is sufficient. Moreover, the website's description, part of the \texttt{meta} tag is part of Resource Description Framework (RDF) that can be used for cataloging~\cite{candan2001resource}. Consequently, we truncated the website content to the first $1,000$ characters and used that for translation. The translator was set up to automatically detect the source language and translate it into English. We successfully translated content from $7,399$ websites.

\subsubsection{Website Categorization}
\label{sec:categorization}
We performed the categorization using Google NLP API using the the \texttt{google-cloud-language 2.9.0} Python module~\cite{google_nlp}. Prior works have extensively used Google NLP for sentiment analysis~\cite{shalkarbayuli2018comparison}. We make use of the content classification feature of this API, as previously used to detect fake news~\cite{ibrishimova2020machine} and create knowledge graphs~\cite{lukasik2018content}. Using this API requires a Google Cloud Developer account with the Natural Language API enabled for categorization. The response classifies websites into one to four levels of categories and sub-categories. For example, some websites only received one top-level category like Adult whereas some websites received up to four-levels of categories like ``Business \& Industrial $\rightarrow$ Shipping \& Logistics $\rightarrow$ Freight Transport $\rightarrow$ Maritime Transport." Along with this, we also captured the confidence level as percentage of the NLP classifier for each website. At this stage, we obtained classification for a total of $6,914$ websites. From these, we filtered websites using our inclusion criteria. We selected websites classified with at least $75\%$ confidence level similar to prior work by Kumar et al.~\cite{kumar2023towards} obtaining $3,200$ websites.

To verify the categorization method, we first determined the minimum number of websites required for manual verification. We calculated the z-score using a $95\%$ confidence level and a $10\%$ margin of error for our dataset of $3,200$ websites, resulting in a sample size of $94$. Consequently, we selected a subset of $100$ websites, exceeding the minimum required size, by generating unique random numbers using the python~\texttt{random} library. We sorted the websites by name and used these random numbers as indices to select $100$ websites. To manually identify these websites' categories, we used a mix of understanding the website's category through researchers knowledge, checking the ``about" page of the website, and using Symantec URL categorization tool~\footnote{\url{https://sitereview.bluecoat.com/}}. For websites belonging to multiple categories, we used the primary category of the website and matched that with the Google NLP category with the highest confidence level. Through our manual verification process, we obtained a $91\%$ accuracy score for categorization of websites with at least $75\%$ confidence level. 

\begin{figure*}[ht!]
\centering
\includegraphics[width=1.1\textwidth]{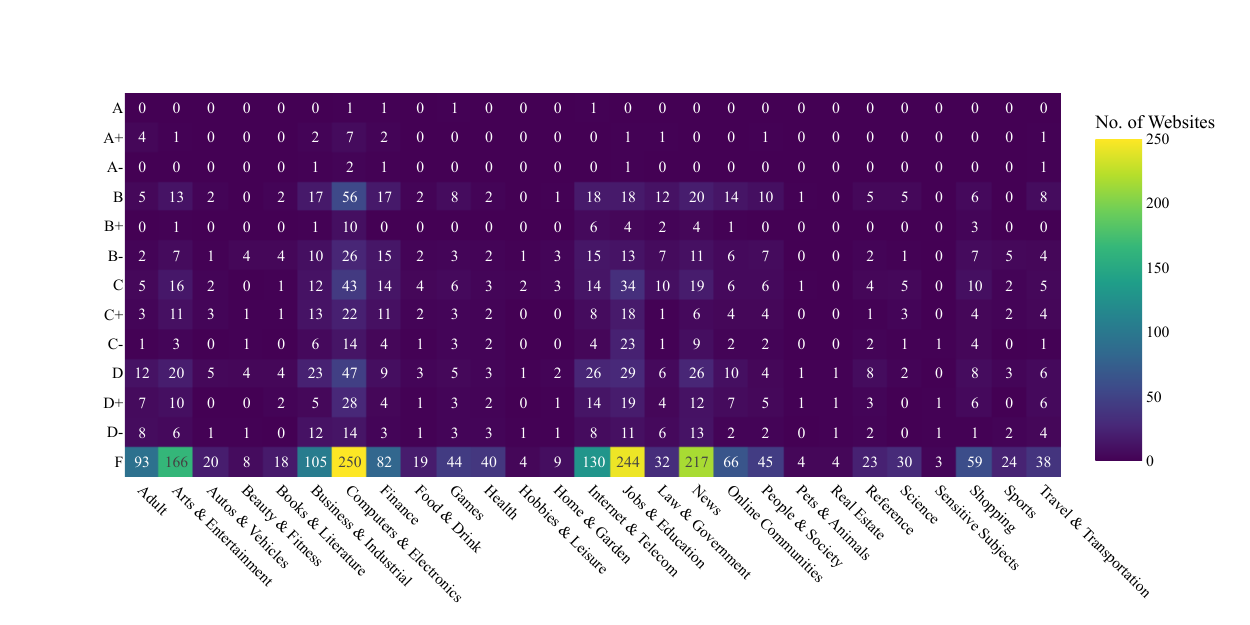}
\caption{Distribution of Websites Between Different Grades and Website Categories}
\label{fig:grade_heatmap}
\end{figure*}

\subsection{Evaluation using Mozilla Observatory}
\label{sec:observatory}
We utilized the Mozilla Observatory tool developed by Mozilla, which automatically analyzes and reports on different HTTP security headers based on Mozilla's web security guidelines~\cite{mozilla_guidelines} which in turn are based on OWASP compliance~\cite{owaspWSTGLatest}. This tool performs a series of tests on the HTTP related headers of the websites, to determine the adoption of HTTPS for that website~\cite{felt2017measuring}. The tool reports various security measures for $12$ different categories, with each category having an associated security benefit and implementation difficulty tagged with either Maximum, High, Medium, and Low. For example, for cookies category, it checks whether the cookies are secured with the HttpOnly attribute and whether any cookies were found in the first place. Each website begins with a score of $100$ and then receives a final score based on the scores from each category. Depending on the analysis, each category receives either a positive score, a negative score, or a base score of zero. This modifier is always a multiple of five, so the final score is always a multiple of five. A positive score indicates that the website has taken extra measures to strengthen the security for that header. A negative score indicates that the baseline implementation for the security header is not met, which may leave the website open to vulnerabilities~\cite{mozilla_risk}.

Most categories give the websites either a score of zero or a negative score. If the website scores $90$ or more, then it can receive extra credit from six possible categories: Cookies, Content Security Policy, HTTP Strict Transport Security, Subresource Integrity, and X-Frame-Options. The maximum possible score a website can receive is $135$, and the minimum possible score is always zero. If a website's score aggregates to be less than zero, then it is capped at zero. The reported categories and their possible min-max score pairs are: Content Security Policy ($+10$, $-25$), Cookies ($+5$, $-40$), Cross-origin Resource Sharing ($0$, $-50$), Public Key Pinning ($0$, $-5$), Redirection from HTTP ($0$, $-20$), Referrer Policy ($+5$, $-5$), Strict Transport Security ($+5$, $-20$), Subresource Integrity ($+5$, $-50$), X-Content-Type-Options ($0$, $-5$), X-Frame-Options ($+5$, $-20$), X-XSS-Protection ($0$, $-10$), and configuration of Contribute JSON ($0$, $-10$). 

Each score range has an associated letter grade given to the website. The possible grades along with their score ranges (in multiples of five) are: A+ ($100-135$), A ($90-95$), A- ($85$), B+ ($80$), B ($70-75$), B- ($65$), C+ ($60$), C ($50-55$), C- ($45$), D+ ($40$), D ($30-35$), D- ($25$), and F ($0-20$). The full report provides details of each category, the score modifier, and reason for the score. The tabular report offers an overview of the score in each category, the final score, and the letter grade. We note that the cookie analysis here does not give a count of how many cookies on the website were analyzed. We only obtain an indication of whether there were no cookies or at least one cookie. On a Linux virtual machine, we created a Bash script to automatically run all $3,200$ websites through this tool. We collected reports successfully for $3,195$ websites, which is sufficient for website analysis as illustrated by prior work using $1,000$ websites~\cite{dewald2010adsandbox}. Only five websites did not result in a successful run with the tool.

\begin{table*}[p]
\caption{Website Categories Along with their Top Three Sub-Categories List, Count of Websites (\#), Average (Avg), and Maximum (Max) Scores (The Minimum Score for Each Category is Zero)}
\centering
\renewcommand{\arraystretch}{0.9} % Adjust row height (reduce vertical space)
\resizebox{\textwidth}{!}{
\begin{tabular}{lp{5.5cm}p{1cm}p{1cm}p{1cm}p{1cm}}
\toprule
\textbf{Category} & \textbf{Sub-Categories} & \textbf{\#} & \textbf{Avg} & \textbf{Max}\\ \midrule
Computers \&   Electronics            & Software, Enterprise Technology, Computer   Security                      & 520                                & 31.97                                                             & 125                              \\
Jobs \& Education                     & Education, Jobs, Internships                                              & 415                                & 23.78                                                             & 115                              \\
News                                  & Other, Technology News, Sports News                                       & 337                                & 20.21                                                             & 80                               \\
Arts \& Entertainment                 & TV \& Video, Music \& Audio, Comics \& Animation                          & 254                                & 19.94                                                             & 100                              \\
Internet \& Telecom                   & Web Services, Mobile \& Wireless, Email \& Messaging                      & 244                                & 27.87                                                             & 95                               \\
Business \& Industrial                & Business Services, Advertising \& Marketing, Business Operations          & 207                                & 28.41                                                             & 110                              \\
Finance                               & Banking, Investing, Insurance, Credit \& Lending                          & 163                                & 32.30                                                             & 115                              \\
Adult                                 &                                                                           & 140                                & 18.75                                                             & 105                              \\
Online Communities                    & File Sharing \& Hosting, Social Networks, Photo \& Video Sharing          & 118                                & 26.02                                                             & 80                               \\
Shopping                              & Shopping Portals, Classifieds, Consumer Resources                         & 108                                & 27.64                                                             & 80                               \\
People \& Society                     & Social Issues \& Advocacy, Religion \& Belief, Other                      & 86                                 & 29.07                                                             & 100                              \\
Law \& Government                     & Government, Legal, Military                                               & 82                                 & 35.67                                                             & 110                              \\
Games                                 & Computer \& Video Games, Roleplaying Games, Gambling                      & 79                                 & 26.46                                                             & 95                               \\
Travel \& Transportation              & Transportation, Hotels \& Accommodations, Travel Guides \&   Travelogues  & 78                                 & 29.49                                                             & 110                              \\
Health                                & Medical Facilities \& Services, Health Conditions, Mental Health          & 59                                 & 18.14                                                             & 75                               \\
Reference                             & General Reference, Language Resources, Geographic Reference               & 50                                 & 28.50                                                             & 75                               \\
Science                               & Other, Computer Science, Mathematics                                      & 47                                 & 24.89                                                             & 75                               \\
Sports                                & Team Sports, College Sports, Individual Sports                            & 38                                 & 20.13                                                             & 65                               \\
Food \& Drink                         & Cooking \& Recipes, Food \& Grocery Delivery, Food \& Grocery   Retailers & 35                                 & 27.14                                                             & 75                               \\
Autos \& Vehicles                     & Motor Vehicles (By Brand), Vehicle Shopping, Motor Vehicles (By Type)     & 34                                 & 24.12                                                             & 75                               \\
Books \& Literature                   & E-Books, Other, Audiobooks                                                & 32                                 & 26.41                                                             & 75                               \\
Home \& Garden                        & Home Improvement, Home \& Interior Decor, Home Safety \& Security         & 20                                 & 30.75                                                             & 75                               \\
Beauty \& Fitness                     & Fashion \& Style, Face \& Body Care, Fitness, Beauty Services \&   Spas   & 19                                 & 29.21                                                             & 65                               \\
Hobbies \& Leisure                    & Special Occasions, Crafts, Outdoors                                       & 9                                  & 26.67                                                             & 65                               \\
Pets \& Animals                       & Animal Products \& Services, Wildlife                                     & 8                                  & 25.63                                                             & 70                               \\
Real Estate                           & Real Estate Services, Real Estate Listings                                & 7                                  & 21.43                                                             & 40                               \\
Sensitive Subjects                    & Death \& Tragedy, Recreational Drugs, War \& Conflict                     & 6                                  & 21.67                                                             & 45                               \\
\bottomrule
\end{tabular}
}
\label{tab:category}
\end{table*}

\section{Results}
\label{sec:results}

\subsection{Classification \& Distribution of Categories}
We classified the websites into various levels of categories and sub-categories chains. The classification yielded a total of $27$ top-level categories (e.g., Shopping), $191$ two-level category chains (e.g., Arts \& Entertainment $\rightarrow$ Celebrities \& Entertainment News), $433$ three-level category chains (e.g., Autos \& Vehicles $\rightarrow$ Vehicle Codes \& Driving Laws $\rightarrow$ Vehicle Licensing \& Registration), and $433$ four-level category chains (e.g., Business \& Industrial $\rightarrow$ Shipping \& Logistics $\rightarrow$ Freight Transport $\rightarrow$ Maritime Transport). Only three categories had complete four-level category chains, with each leaf node being unique, resulting in the same number for both three-level and four-level category chains. Table~\ref{tab:category} displays the top three sub-categories within each category.

\subsection{Distribution of Security Grades and Scores}
Mozilla Observatory assigns a letter-grade and a score to each website. More than half of the websites ($n=1,777$, $55.62$\%) received an F grade, including $32.71\%$ ($n=1,045$) that scored zero, as shown in Figure~\ref{fig:grade_heatmap}. We observed peaks at grades A+, B, C, and D, which encompass a broader range of scores. Figure~\ref{fig:grade_heatmap} illustrates the distribution of website counts for both grades and website categories. The overall average score across all websites is $26.21$. Upon examining the category-wise distribution of scores in Table~\ref{tab:category}, we find that the top five categories with the highest average scores are Law \& Government, Finance, Computers \& Electronics, Home \& Garden, and Travel \& Transportation. The bottom five categories with the lowest average scores are Health, Adult, Arts \& Entertainment, Sports, and News. The highest score of $125$ was achieved by only one website, \textit{\url{www.gimp.org}}, a free and open-source image editor in the Computers \& Electronics category. Further, we note that despite their purpose, websites in the Computer Security category ($n=66$) scored only an average of $31.89$, with min-max score pair ($0$,$105$).

\subsection{Analysis of HTTP Security Headers}
\begin{figure*}[htp!]
\centering
\includegraphics[width=1.1\textwidth]{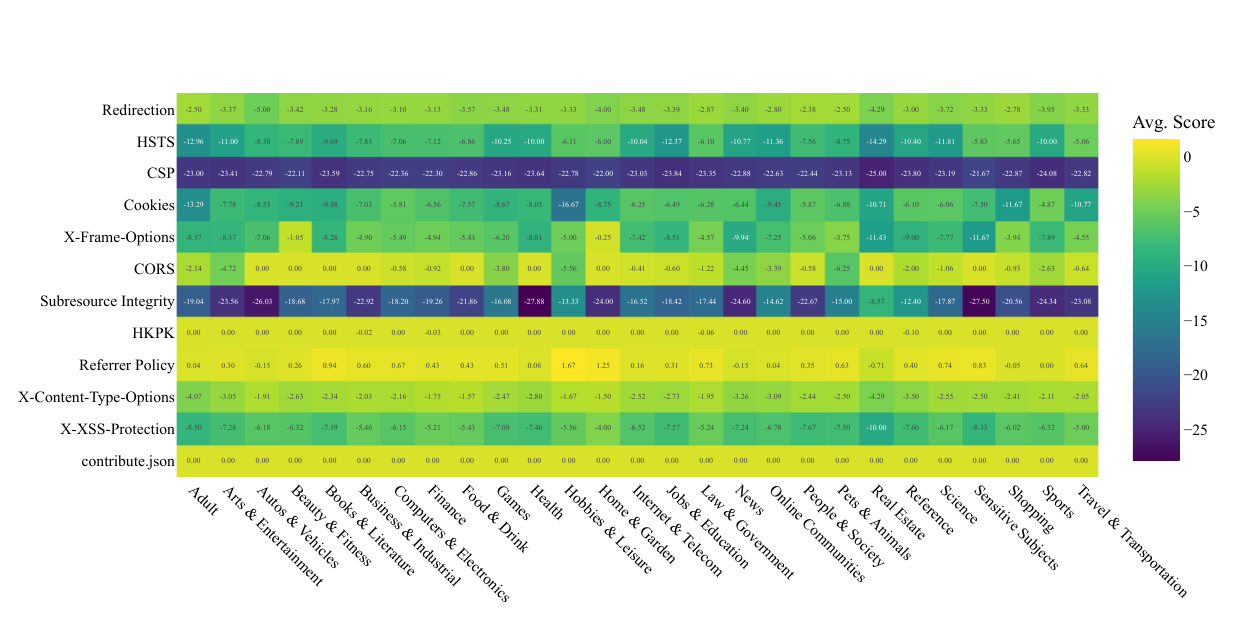}
\caption{Heatmap of the Distribution of Average HTTP Security Score between Different Security Headers and Website Categories}
\label{fig:website_metric_heatmap}
\end{figure*}

Mozilla Observatory provides a report on $12$ security measures for each website, each having an associated security benefit: Maximum, High, Medium, and Low. We analyze the results based on the security benefit provided by each measure to offer guidance to researchers and developers prioritizing the most relevant measures. Figure~\ref{fig:website_metric_heatmap} presents a heatmap of the scores obtained across each website category per security measure. 

\subsubsection{Maximum Benefit}
When a user or browser connects to a website using the Hypertext Transfer Protocol (HTTP), the connection is typically served at port 80 of the website's server. The security standard requires websites to use Hypertext Transfer Protocol Secure (HTTPS) instead, typically served at port $443$. Since users often type website URLs starting with either \texttt{http://} or \texttt{https://}, website servers usually listen on both ports and must securely redirect HTTP traffic to HTTPS. This security measure is considered to have low implementation difficulty by Mozilla. All website categories received an average negative score for redirection. The top five categories with the highest average scores are: People \& Society, Adult, Pets \& Animals, Shopping, and Online Communities. In contrast, the bottom five categories with the lowest average scores are: Autos \& Vehicles, Real Estate, Home \& Garden, Sports, and Science. 

\subsubsection{High Benefit}
\label{sec:high_benefit}
Five of the categories by Mozilla Observatory were identified as a High security benefit and are discussed below. Addressing metrics with high security benefits can close significant HTTP implementation security gaps.

\textbf{Strict Transport Security:}
Unlike redirection, where users visiting a site using the \texttt{http://} protocol are redirected to the website using \texttt{https://} by the website's server, HTTP Strict Transport Security (HSTS), an HTTP header, enables the browser to perform this action. After the server's first redirection, the browser may ``remember" the protocol change and automatically upgrade all future requests by transparently changing the protocol in the URL. HSTS enhances website security by preventing users from bypassing any Transport Layer Security (TLS) and certificate-related errors. Weak HSTS implementation can make websites vulnerable to attacks such as SSLStrip~\cite{selvi2014bypassing}. Mozilla identifies this security measure as having low implementation difficulty. All website categories received an average negative score for Strict Transport Security. The top five categories with the highest average scores include Travel \& Transportation, Shopping, Sensitive Subjects, Home \& Garden, and Law \& Government. In contrast, the bottom five categories with the lowest average scores are Real Estate, Adult, Jobs \& Education, Science, and Online Communities.

\textbf{Content Security Policy:}
Websites often use resources such as JavaScript libraries, Cascading Style Sheets (CSS), and media like images and videos, which can be loaded from external sources. In some cases, external sources may serve malicious scripts that lead to XSS attacks. The \texttt{Content-Security-Policy} (CSP) is an HTTP header that allows developers to control which external sources are used for a website's resources. The CSP header also provides additional protection against XSS attacks by disabling inline JavaScript. Mozilla identifies this security measure as having high implementation difficulty. All website categories received an average negative score for Content Security Policy. The top five categories with the highest average scores include Sensitive Subjects, Home \& Garden, Beauty \& Fitness, Finance, and Computers \& Electronics. The bottom five categories with the lowest average scores are Real Estate, Sports, Jobs \& Education, Reference, and Health.

\textbf{Cross-Origin Resource Sharing: }
Cross-Origin Resource Sharing (CORS) helps website specify to the browser which external resources are permitted for loading resources from the website. The external resources can be identified either by the domain name, the protocol or scheme, or the port. For this, the HTTP header \texttt{Access-Control-Allow-Origin} can be used to specify the allowed external sources. This security measure is identified with low implementation difficulty by Mozilla. Nine website categories received an average score of zero whereas the remaining website categories received average negative score for CORS. The nine categories that received an average score of zero are: Home \& Garden, Autos \& Vehicles, Sensitive Subjects, Beauty \& Fitness, Books \& Literature, Real Estate, Business \& Industrial, Food \& Drink, and Health. The bottom five categories with the lowest average scores are: Pets \& Animals, Hobbies \& Leisure, Arts \& Entertainment, News, and Games. 

\textbf{Cookies: }
Due to the stateless nature of the HTTP and HTTPS protocols, the connection between a client (web browser) and a website's server does not maintain any state-related data. Cookies, small blocks of data storing information on the browser, provide a means for maintaining state between the client and the server~\cite{park2000secure}. Since cookies may contain sensitive user account information and other identifying details, it is crucial to limit cookie access to the relevant websites or domains, thereby mitigating attacks such as session hijacking~\cite{dacosta2012one}. Mozilla identifies this security measure as having medium implementation difficulty. All website categories received an average negative score for Cookies. The top five categories with the highest average scores are: Sports, Computers \& Electronics, People \& Society, Science, and Reference. On the other hand, the bottom five categories with the lowest average scores are: Hobbies \& Leisure, Adult, Shopping, Travel \& Transportation, and Real Estate.

\textbf{X-Frame-Options: }
Clickjacking is a common type of website-based attack, where a user is deceived into clicking a hidden or disguised element belonging to a different page, while assuming they are interacting with something else on their current page~\cite{huang2012clickjacking}. It is the developer's responsibility to ensure that their website cannot be used in clickjacking attacks originating from other websites. The \texttt{X-Frame-Options} (XFO) HTTP header can be employed to control how a website is used within an Inline Frame or iFrame, an HTML element that can load an HTML page from within another HTML page. Effective implementation of both XFO and CSP can help protect against clickjacking~\cite{calzavara2020tale}. Mozilla identifies this security measure as having low implementation difficulty. All website categories received an average negative score for X-Frame-Options. The top five categories with the highest average scores are: Home \& Garden, Beauty \& Fitness, Pets \& Animals, Shopping, and Travel \& Transportation. Conversely, the bottom five categories with the lowest average scores are: Sensitive Subjects, Real Estate, News, Reference, and Health. 
\subsubsection{Medium Benefit}
Only one category reported by the Mozilla Observatory was identified as having a medium security benefit: Subresource Integrity.

\textbf{Subresource Integrity: }
When websites load external JavaScript libraries from public content delivery networks (CDNs) like JQuery \textit{\url{jquery.org}}, they become vulnerable to attacks if these external libraries get corrupted or modified due to an attack on the CDN itself. Such an attack on the CDN can lead to further attacks, such as denial of service (DoS) or credential theft, on all websites using the external library. To protect against these attacks, the World Wide Web Consortium (W3C) introduced the Subresource Integrity (SRI) standard. SRI allows external resources to be identified with their version, enabling the detection of modifications and preventing the loading of altered resources. Mozilla identifies this security measure as having medium implementation difficulty. All website categories received an average negative score for Subresource Integrity. The top five categories with the highest average scores are: Real Estate, Reference, Hobbies \& Leisure, Online Communities, and Pets \& Animals. The bottom five categories with the lowest average scores are: Health, Sensitive Subjects, Autos \& Vehicles, News, and Sports. 
\subsubsection{Low Benefit}
Mozilla Observatory identified five categories as providing low security benefits, which are discussed below.

\textbf{Public Key Pinning: }
HTTP Public Key Pinning (HPKP) enables websites to associate their site with specific endpoint public keys, intermediate certificate authorities, or root certificate authorities. This prevents attackers from tricking certificate authorities into issuing unauthorized certificates for websites, reducing the risk of Man-in-the-Middle (MitM) attacks. Mozilla classifies this security measure as having maximum implementation difficulty. Four website categories received an average negative score for Public Key Pinning and the remaining received an average score of zero. The website categories that received an average negative scores (bottom four categories) are: Reference, Law \& Government, Finance, and Business \& Industrial.

\textbf{Referrer Policy: }
The HTTP Referrer Policy enables developers to control how external requests are handled, minimizing the risk to user privacy. Mozilla identifies this security measure as having low implementation difficulty. Four website categories received an average negative score for Referrer Policy, one category received an average zero score, and the remaining categories received positive scores. The top five categories with the highest average scores are: Hobbies \& Leisure, Home \& Garden, Books \& Literature, Sensitive Subjects, and Science, whereas, the bottom five categories with the lowest average scores are: Real Estate, News, Autos \& Vehicles, Shopping, and Sports.

\textbf{X-Content-Type-Options: }
\texttt{X-Content-Type-Options} HTML header specifies that browsers should only load external scripts and stylesheets if they are tagged with the correct Multipurpose Internet Mail Extensions (MIME) type. This prevents malicious content from being loaded onto websites, protecting against XSS attacks. Mozilla identifies this security measure as having low implementation difficulty. All website categories received an average negative score for X-Content-Type-Options. The top five categories with the highest average scores are: Home \& Garden, Food \& Drink, Hobbies \& Leisure, Finance, and Autos \& Vehicles, whereas, the bottom five categories with the lowest average scores are: Real Estate, Adult, Reference, News, and Online Communities.

\textbf{X-XSS-Protection: }
The X-XSS-Protection HTTP header prevents XSS attacks in older browsers that do not support Content Security Policy (CSP). Mozilla identifies this security measure as having low implementation difficulty. All website categories received an average negative score for X-XSS-Protection. The top five categories with the highest average scores are: Home \& Garden, Travel \& Transportation, Finance, Law \& Government, and Food \& Drink, whereas, the bottom five categories with the lowest average scores are: Real Estate, Adult, Sensitive Subjects, People \& Society, and Reference. 

\section{Discussion and Implications}
\label{sec:discussions}
The overall security of popular websites is alarmingly inadequate, prompting concerns about the potential risks to user data and the general online experience. A total of 1,777 ($55.62\%$) websites received a grade `F', with each category containing at least one website with a score of zero. In fact, $1,045$ ($32.71\%$) websites received a score of zero, and the average overall score was a meager $26.21$ out of a possible $135$. Given that these are some of the most popular websites, our findings emphasize the urgency for substantial improvements in their security practices. As browser security indicators such as the security padlock next to URL are not entirely trustworthy and may sometimes indicate about a website's HTTP implementation strength erroneously~\cite{felt2016rethinking}. The adoption of Mixed Content policy~\footnote{\url{https://w3c.github.io/webappsec-mixed-content/}} have ensured blocking non-secure content from being loaded in the browser, however, this measure still lacks in mobile device browsers~\cite{chen2015dangerous}. This can arise due to inconsistencies in HTTP header implementation across desktop and mobile devices~\cite{mendoza2018uncovering}. Furthermore, as seen in Figure~\ref{fig:website_metric_heatmap} metrics of CSP and SRI have received the lowest possible scores overall. 

\subsection{Critical Websites}
\label{subsec:health_website_security}
Our category-wise analysis revealed that health-related websites had the lowest security scores. This is concerning because health information is sensitive and subject to strict regulations in many countries~\cite{baker2006privacy}. This could put patients' data at risk and may result in noncompliance with data protection laws such as Health Insurance Portability and Accountability Act (HIPAA) in the United States~\cite{act1996health}, Personal Data Protection Bill (PDPB) in India~\cite{prasad2020personal}, Personal Information Protection Act (PIPA) in South Korea~\cite{ko2017structure}, and Personal Information Protection and Electronic Documents Act (PIPEDA) in Canada~\cite{jaar2008canadian}. Further research is needed to understand the specific factors that contribute to the poor security performance of health websites, which may include inadequate technical expertise, resource constraints, or insufficient prioritization of security~\cite{harvey2014privacy}. Another concerning category that received a low security score is adult websites. Poor security on adult websites can lead to significant risks, including the exposure of users' personal and sensitive data, and increased vulnerability to malware and phishing attacks~\cite{vallina2018my}. Leakage of personal information through these attacks on adult websites can cause psychological distress for users and result in reputation damage for the websites involved~\cite{vallina2019tales}. 

\subsection{Secure Redirection}
\label{subsec:redirection}
Improper redirection can lead to Session Hijacking attacks~\cite{cheng2010analysis}. Our analysis showed that all categories had an average negative score for the redirection metric. Only 53.02\% of websites implemented redirection to HTTPS correctly, while the rest either lacked a secure redirection mechanism or had suboptimal implementations. Investigating the barriers to adoption of secure redirection, such as developer awareness or misconceptions about implementation complexity, could help identify strategies for encouraging more widespread implementation of this crucial security measure. For example, developers might underestimate the importance of secure redirection, assuming that other security measures are sufficient, or they might be unaware of the latest best practices and tools that facilitate secure implementation. The implementation of strict redirection policies, HST), and CSP in conjunction can achieve robust secure redirection~\cite{chang2017security}. 

\subsection{Benefits with Implementation}
% \label{subsec:high_benefits_low_difficulty}
Despite the high security benefits and low implementation difficulty of HSTS, X-Frame-Options, and CORS, websites continue to underperform in these categories. All website categories had negative average scores for HSTS and X-Frame-Options, while $18$ of the $27$ categories had negative average scores for CORS. These low scores resulted from bad implementation practices including short HSTS validity, not implementing X-Frame-Options, and having visible content via CORS headers or files. Prior work shows that bad practices and incorrect implementation of HSTS and HPKP can lead to further security vulnerabilities~\cite{de2016implementation}. To improve the security of websites, developers should prioritize these metrics~\cite{dolnak2017introduction}. Research into the reasons behind this underperformance, such as a lack of awareness, technical challenges, or insufficient motivation, could help inform targeted interventions and resources to support developers in implementing these critical security measures.

\section{Limitations and Future Work}
\label{sec:limitations}
We analyzed $N=3,195$ websites selected from the top $M=10,000$ websites on the Tranco list. In the future, we plan to scale up our analysis to encompass all $1M$ websites on Tranco's list. We also focused on the HTTP security headers of websites using Mozilla Observatory. Although Mozilla Observatory offers valuable insights into a website's implementation of $12$ important HTTP headers, the tool could not successfully run on $5$ out of the $3,200$ categorized websites. Although this study was focused on analyzing HTTP security headers, in future, we aim to explore additional security and privacy aspects, including cookie variants, compliance with privacy regulations, data storage and handling protocols, and user authentication methodologies.

\section{Conclusion}
\label{sec:conclusion}
We conducted security evaluation of popular websites, focusing on HTTP security headers across $27$ website categories—a largely understudied domain. We analyzed the top $M=10,000$ globally popular websites, employing a three-step process of website content scraping, content translation, and NLP-based categorization to determine website categories. Subsequently, Mozilla Observatory was used to perform a detailed evaluation of HTTP security headers on $N=3,195$ websites. The results revealed significant security deficiencies: $55.66\%$ of the websites received a failing grade (`F'), and $32.71\%$ scored zero out of a possible $135$ points. Healthcare websites were particularly concerning, ranking lowest in security scores, highlighting risks to sensitive health-related data and potential non-compliance with stringent healthcare regulations. Secure redirection from HTTP to HTTPS, despite its critical importance and relatively low implementation complexity, was poorly handled across many sites. The study emphasizes the need for a systematic approach to implementing and auditing security headers to mitigate potential threats and secure the web infrastructure effectively.

\section{Acknowledgement}
We would like to acknowledge the Inclusive Security and Privacy-focused
Innovative Research in Information Technology (InSPIRIT) Lab for supporting this work. Any opinions, findings, conclusions, or recommendations expressed in this material are
solely those of the author.

\bibliographystyle{splncs04}
\bibliography{ICISS}

\end{document}